\documentclass{icrc2009}

\usepackage{graphicx}   
\usepackage{caption}    
\usepackage[font=footnotesize]{subfig} 
\usepackage{fixltx2e}
\usepackage{url}

\newcommand{\shorttitle}[1]%
{\markboth{Proceedings of the 31\MakeLowercase{$^{st}$} ICRC, {\L}\'{o}d\'{z} 2009}{#1} }
\newcommand{\etal}{\MakeLowercase{\textit{et al. }}} 

\usepackage{cite}

\def\la{\mathrel{\mathchoice {\vcenter{\offinterlineskip\halign{\hfil
$\displaystyle##$\hfil\cr<\cr\sim\cr}}}
{\vcenter{\offinterlineskip\halign{\hfil$\textstyle##$\hfil\cr
<\cr\sim\cr}}}
{\vcenter{\offinterlineskip\halign{\hfil$\scriptstyle##$\hfil\cr
<\cr\sim\cr}}}
{\vcenter{\offinterlineskip\halign{\hfil$\scriptscriptstyle##$\hfil\cr
<\cr\sim\cr}}}}}

\def\apj{ApJ}%
\def\apjl{ApJ}%
\def\aap{A\&A}%
\def\mnras{MNRAS}%
\def\prd{Phys.~Rev.~D}%
\begin{document}
\title{H.E.S.S. VHE gamma-ray observations of the microquasar\\ GRS 1915+105}

\author{\IEEEauthorblockN{A. Szostek\IEEEauthorrefmark{1}\IEEEauthorrefmark{2},
G. Dubus\IEEEauthorrefmark{1}, F. Brun\IEEEauthorrefmark{3} and M. de
Naurois\IEEEauthorrefmark{3} for H.E.S.S. Collaboration}\\
\IEEEauthorblockA{\IEEEauthorrefmark{1}Laboratoire d'Astrophysique de Grenoble, INSU/CNRS, Universit\'e Joseph Fourier,\\ BP 53, F-38041 Grenoble Cedex 9, France}
\IEEEauthorblockA{\IEEEauthorrefmark{2}Obserwatorium Astronomiczne, Uniwersytet
Jagiello{\'n}ski, ul. Orla 171, 30-244 Krak{\'o}w, Poland}
\IEEEauthorblockA{\IEEEauthorrefmark{3}LPNHE, Universit\'e Pierre et Marie
Curie Paris 6, Universit\'e Denis Diderot Paris 7,\\ CNRS/IN2P3, 4 Place
Jussieu, F-75252, Paris Cedex 5, France}
}

\shorttitle{A. Szostek \etal GRS 1915+105}
\maketitle

\begin{abstract}
GRS 1915+105 is a very well studied microquasar with a wide variety of temporal and spectral states. Multi-wavelength observations from radio to X-rays have uncovered clear relations between the variability in different bands. GRS 1915+105 is also well-known as a source of superluminal jets moving away from the core with true velocity $\ge 0.9$c. Non-thermal emission from the jets or their termination shock could extend to the VHE gamma-ray domain. GRS 1915+105 was observed with the H.E.S.S. telescope array between 2004 and 2008 for a total of 24 hours. We report on these observations and discuss our findings.
\end{abstract}

 
\section{Introduction}
Microquasars are X-ray binaries which contain relativistic jets which are detected and resolved thanks to their non-thermal radio emission. The canonical example of a microquasar is GRS 1915+105. Its multi-wavelength observations from radio to X-rays have shown a very complex behavior. In particular, a rich phenomenology observed in the radio domain includes radio quiet states, prolonged phases of relatively bright emission from a compact jet, the strong radio flares associated with major ejections in the superluminal jets and rapid radio oscillations. 

Strong observational evidence exists that superluminal jets in Galactic sources are places where particles can be accelerated to very high energies (VHE) \cite{corbel2002,corbel2005}.  If such acceleration takes place in the jet of GRS 1915+105, \cite{atoyan1999} predicted that GRS 1915+105 could emit VHE gamma-rays for a few days, starting from an ejection event traced by a strong radio flare. The estimated power of the jets in GRS 1915+105 exceeds $10^{38}$ ergs s$^{-1}$ \cite{fender1999}. Only a small fraction of this jet power needs to be converted to VHE gamma-rays to produce a detectable signal. 

A detection of VHE emission from another microquasar, Cyg X-1, was reported by the MAGIC collaboration. It occurred during a serendipitous bright X-ray flare that lasted several days \cite{magic-cygx1}. 

Here we report and discuss H.E.S.S. observations of GRS 1915+105.

\section{H.E.S.S. data and analysis}

H.E.S.S. is an array of four imaging atmospheric Cherenkov telescopes situated in the Khomas Highland of Namibia. GRS 1915+105 was observed between year 2004 and 2008 on several occasions, as a target of opportunity (ToO) or as a field of view source in other H.E.S.S. pointings. After applying the standard H.E.S.S. data quality selection criteria \cite{hess-crab} a total of 24.1 hours live time were available for the analysis. The zenith angle of the observations changed between $33^{\circ}$ and $52^{\circ}$ with a mean value of $Z_{\rm mean} = 37^{\circ}$. The pointing offsets varied from $0.45^{\circ}$ to $1.79^{\circ}$. In accordance with H.E.S.S. guidelines, all the results presented below have been successfully cross-checked with an independent analysis and calibration chain.

We performed a point source analysis at the location of the source using the standard analysis techniques and selection cuts with an angular cut of $0.0125$ and a size cut of 80 photo-electrons \cite{hess-calibration, hess-pks2155}. The post-analysis energy threshold at $Z_{\rm mean}$ is 410 GeV. To estimate the background we used the Reflected Background method for all the observational runs \cite{hess-crab}. The total number of source and background events is $1052$ and $15790$ respectively (with background to source normalization equal 0.067), which results in an excess of -5.93 events. The significance of the excess from the direction of GRS 1915+105, calculated using Eq. (17) of \cite{liandma}, is $-0.27$ standard deviations. 

\begin{figure}
\centering
\includegraphics[width=2.5in]{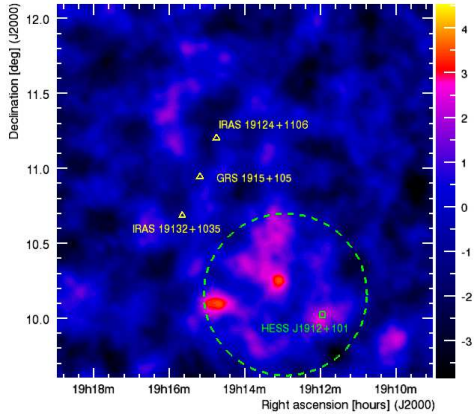}
\caption{Smoothed significance map of region around GRS 1915+105. The integration radius is 0.11$^{\circ}$ and the map has been smoothed with a two-dimensional Gaussian of radius 0.65$^{\circ}$. The triangles denote GRS 1915+105 and the two IRAS sources, candidates for jet-ISM interaction sites. The dotted circle corresponds to an exclusion region around HESS J1912+101. The data inside not taken into account in background calculations.}
\label{map}
\end{figure}

\begin{figure}
\centering
\includegraphics[width=2.6in]{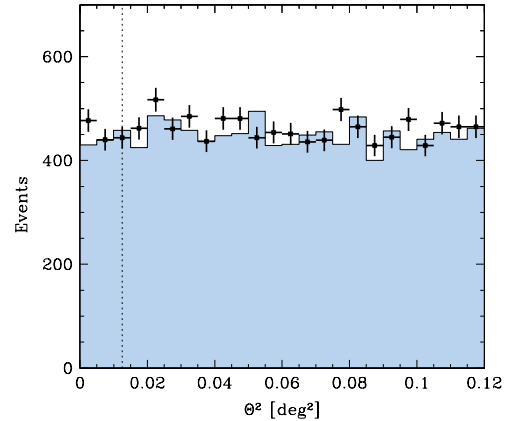}
\caption{Distribution of events as a function of the squared angular distance from GRS1915+105 for gamma-ray-like events in the ON region (points) and from the center of the OFF regions (filled area). The vertical dotted line denotes the standard selection cut for point sources used by H.E.S.S.}
\label{thetaplot}
\end{figure}

Fig. \ref{map} shows a significance map around the position of GRS 1915+105 and Fig. \ref{thetaplot} the distribution of the squared angular distance of observed gamma-ray candidates from the center of the binary in comparison to background data. The angular distribution of the source region events is compatible with the distribution of the background regions events. We conclude, that there is no evidence for a VHE gamma-ray signal from the microquasar GRS 1915+105.

We used approach of \cite{feldman1998} to calculate the upper limits on the integrated photon flux above 410 GeV. Assuming a power law with photon index $\Gamma = 2.5$, the upper limit for the whole data set at the 99.9\% confidence level is
\begin{equation}
I (>410 {\rm\ GeV}) < 0.61 \times 10^{-12} \quad {\rm cm}^{-2} {\rm s}^{-1}.
\end{equation}
This corresponds to 0.71\% of the Crab Nebula flux with the same energy threshold. The estimated systematic error on the flux measurements is 20\% \cite{hess-crab}. The H.E.S.S. upper limits translate into a VHE luminosity $\la10^{34}$ erg\ s$^{-1}$ at the 9 kpc distance for the total dataset and $\la10^{35}$ erg\ s$^{-1}$ for a 4$\sigma$ detection in one hour.

\section{Discussion and conclusions}

In GRS 1915+105, both discrete relativistic ejections and a compact jet could be associated with VHE emission. Unlike Cyg X-1 or LS 5039, VHE photon absorption due to pair creation on stellar photons is not expected to be an issue in GRS 1915+105 due to the low luminosity and temperature companion star.

\cite{atoyan1999} modelled the radio flares from GRS 1915+105 as synchrotron emission of expanding plasmons ejected from the system. In the framework of the model, for an assumed acceleration of electrons to 10 TeV and a sub-equipartition magnetic field of 0.05 G, the VHE gamma-ray flux is expected to be at the Crab nebula flux during the first hours of a strong radio outburst, and to subsequently decline to $\la$ 10\% of the Crab flux over a period of a few days. These fluxes would be detectable with the H.E.S.S. telescopes. 

Due to the highly time-dependent character of the VHE emission it would be desirable to have observations simultaneous with one of the large, optically thin radio flares associated with major ejection events but these require long, dedicated campaigns of tens of days to increase chances of catching them. Despite present efforts, these remain difficult to organise with ground-based Cherenkov arrays.  

The VHE emission might be expected on phenomenological grounds from the steady, compact jet. The VHE flare observed in Cyg X-1 \cite{magic-cygx1} occurred when the source was in the hard X-ray power-law state associated with a conical, continuous, self-absorbed compact jet  \cite{blandford1979}. Such a compact jet is also known to be produced in GRS 1915+105 during long plateau states \cite{kleinwolt2002}. However, a flare analogous to that in Cyg X-1, which in 153 minutes of observation time reached 4$\sigma$, could not have been detected. The future Cherenkov Telescope Array with an order of magnitude higher sensitivity will be able to detect such a flare at 9 kpc.

Finally, a possible source of VHE gamma-ray emission are the two small radio and infrared clouds positioned nearly symmetrically with respect to GRS 1915+105 at angular separation of 17' from the system (see Fig. \ref{map}). Because of their position angle being very similar to the position angle of the sub-arcsec radio jets from GRS 1915+105, the clouds have been postulated to be associated with the termination of the jets in the ISM \cite{rodriguez1998}. There is no evidence for an excess VHE gamma-ray emission from any of the two sources. 

A more detailed analysis and discussion in the context of the established variability multi-wavelength behaviour of GRS 1915+105 will be presented in the forthcoming paper.

\section*{Acknowledgment} 
A. Szostek acknowledges support from the European Community via contract ERC-StG-200911 and in part by the Polish MNiSW grant NN203065933


\end{document}